\documentstyle[12pt,epsf]{article}
\begin{document} 
\begin{titlepage}
\begin{flushright}
 JLAB-PHY-99-05
\end{flushright}
\begin{center}
  \bf Pseudoscalar Meson Electroproduction
      Above the Resonance Region\\*[1.0cm]
  \normalsize             Bogdan B. Niczyporuk \\
  \normalsize\sl   Jefferson Lab, Newport News, VA 23606, USA\\*[0.4cm]
  \normalsize\sl   presented \\
  \normalsize\sl   at\\
  \normalsize\sl   CLAS with 6 GeV Beams Workshop, January 12-13, 1999
\end{center}
  
\vskip 2.5cm
\begin{center}
  \bf Abstract
\end{center}
\noindent 
One principal motivation for studying exclusive reactions is that they provide 
a new class of observables, called off-diagonal parton distributions, for the 
internal structure of the nucleon. The study of exclusive reactions provides a 
probe of nucleon structure complementary to purely inclusive studies. The 
simplest, and possibly the most promising, type of experiment is exclusive 
electroproduction of pseudoscalar mesons at small $t$, and at large $Q^2$ and 
$W$. We show that using the CLAS spectrometer at JLAB and with beam energies
between 4 and 6 $GeV$, we can obtain good quality electroproduction data that 
will improve our understanding of nucleon structure.

\end{titlepage}
\newpage
\setcounter{page}{2}

\begin{flushleft}
{\bf 1. Physics Motivations}
\end{flushleft} 
\noindent
The structure of the nucleon revealed in hard processes is described by parton
distributions. Traditionally, the internal structure of the 
nucleon has been studied (CERN, SLAC, DESY) through inclusive scattering 
of high energy leptons in the deep inelastic scattering (DIS) limit, i.e. at 
large $Q^2$, $\nu$, and fixed Bjorken $x = Q^2/2m_p\nu$, where $Q^2$ and 
$\nu$ are the mass squared and energy of the virtual photon. For example, unpolarized
DIS provided the first evidence that quarks carry only about 45$\%$ of the 
nucleon momentum via measurements of the structure function $F_1(x,Q^2)$,
or the corresponding parton densities $q(x,Q^2)$.
Recent measurements have focused on the leading-twist structure function 
$g_1(x,Q^2)$, which is roughly proportional to the inclusive spin asymmetry on a 
longitudinally polarized target. Polarized DIS measurements of $g_1(x,Q^2)$, or 
the relevant helicity densities $\Delta q(x,Q^2)$, have revealed that only about 
30$\%$ of the nucleon's spin is carried by the quark's intrinsic spin \cite{Bass}.

Processes where at least one hadron is detected in the final state offer several 
distinct advantages over inclusive processes alone \cite{NMike}, and references
therein. Particularly interesting is the chirally odd structure function 
$h_1(x,Q^2)$, or the appropriate transversity densities $\delta q(x,Q^2)$ 
\cite{Jaffe,XJi1}. 
Together with $F_1(x,Q^2)$ and $g_1(x,Q^2)$, $h_1(x,Q^2)$ is necessary for a 
complete description of the quark structure of the nucleon in high-energy 
processes. The structure function $h_1(x,Q^2)$ has never been measured.
Chirally odd quark distributions are difficult to measure because they are 
suppressed in totally inclusive deep inelastic scattering. However, the 
asymmetry for semi-inclusive leptoproduction of pions off transversely polarized 
target contains a contribution from $h_1(x,Q^2)$ that is enhanced at low $x$. 
Another motivation for the measurement of $h_1(x,Q^2)$ is a sensitivity to the 
role of relativistic effects in the nucleon state, and a possible sensitivity
\cite{NMike} to gluon contributions to the spin of the proton.

Exclusive electroproduction of mesons from nucleons has become a field of growing 
interest \cite{Vander,Mank,Brod} since a full factorization theorem has been 
proved \cite{Collins,XJi2,Rady}.  
It has been shown that exclusive meson production at large $Q^2$ and small $t$
factorizes into a hard scattering coefficient, a quark-antiquark distribution 
amplitude of the meson, and an off-diagonal quark (or gluon) distribution that 
describes the ``soft" physics in the nucleon. The proof of factorization applies 
when the virtual photon is longitudinally polarized. It has been also shown
\cite{Collins} that transverse polarization of the photon implies a power 
suppression in $Q$ relative to the case of longitudinal polarization. The theorem 
applies to the production of mesons at all $x$. Therefore, off-diagonal (also 
called off-forward \cite{XJi2} or non-forward \cite{Rady}) parton distributions 
allow the description of certain exclusive reactions in the framework of QCD.
 
For longitudinally polarized vector mesons, the relevant parton densities
are the unpolarized ones, $q(x,Q^2)$. For transversely polarized vector mesons, 
the parton densities are the quark transversity densities, $\delta q(x,Q^2)$. 
The original hope \cite{Collins} that $h_1(x,Q^2)$ may be measured via production
of transverse vector mesons has, unfortunately, not come true because in hard 
scattering processes such a transition is forbidden \cite{Hoodb,Diehl}. 
For the pseudoscalar mesons, the relevant parton densities are the quark helicity
densities, $\Delta q(x,Q^2)$, which are not suppressed at large $x$. Hence, the 
polarized parton densities can be probed in unpolarized collisions.

It is clear that the study of exclusive and semi-inclusive reactions provides a 
probe of nucleon structure complementary to purely inclusive studies.
In particular, we should study exclusive reactions at low $t$, and at high $Q^2$ 
and $W$, in the region of validity of QCD factorization theorems, and also in
the transition region where standard partonic model may no longer be valid.
One can then probe in a novel way the soft part of a proton, and elucidate the
transition between soft and hard scattering processes.

A measurement of the differential cross section $\sigma(t,W,Q^2)$ for
the reactions $e^{-} + p \rightarrow e^{-} + \pi^{+}(K^{+}) + n(\Lambda^{o})$
at beam energies 4.0, 4.5, 5.0, 5.5 and 6.0 $GeV$ was proposed and discussed
\cite{Bogdan1}.
Data will be collected simultaneously for $\pi^+$, $\pi^o$ and $K^+$ 
exclusive electroproduction using the CLAS detector at JLAB in the following 
kinematical region: $Q^2 > 1$ $GeV^2$ and $W >2$ $GeV$. In the following sections
existing data, cross section, detailed simulation, reconstruction and analysis of
charged pions electroproduction will be discussed.
 
\begin{flushleft}
{\bf 2. Cross Section}
\end{flushleft}
\noindent
The procedure of extracting a virtual photon cross section $\sigma_{\gamma_v p}$ 
from the observed electroproduction cross section is based on the one-photon 
approximation. In this procedure electrons are regarded as providing a beam of 
virtual photons (flux $\Gamma$) of known polarization $\epsilon$, mass squared 
$Q^2$, and energy $\nu$.  
Electroproduction reactions can be described in terms of form factors
that are generalizations of the form factors observed in elastic
electron-proton scattering, or in terms of cross sections that are extensions
of the photoproduction cross sections.
The most general form of the differential cross section $\sigma$ for the 
reactions
\begin{eqnarray}
  e^{-} + p \rightarrow e^{-} + \pi^{+}(K^{+}) + n(\Lambda^{o})
\end{eqnarray}
can be written in terms of four structure functions (unpolarized data) 
\cite{Bartl}:
\begin{eqnarray}
 \sigma_{\gamma_v p}(W^2,Q^2,t,\phi) & = &
 \sigma_T + \epsilon \sigma_L + \epsilon \sigma_{TT}
 cos2\phi +\sqrt{\epsilon (\epsilon+1)/2}\cdot \sigma_{LT} cos\phi  
\end{eqnarray}
where $\sigma_T$, $\sigma_L$, $\sigma_{TT}$, and $\sigma_{LT}$ 
are functions of the variables: $Q^2$, $W^2 = 2m_p\nu - Q^2 + m_p^2$, and 
$t \equiv (p_{\gamma_v} - p_{\pi,K})^2 - t_{min}$ (or $\theta^*$, the angle 
between the virtual photon and the meson in the hadronic center of mass $W$). 
The dependence on the azimuthal angle $\phi$ (angle of the meson relative to the 
electron scattering plane: $\phi \equiv \phi^*$) is shown explicitly in eq. (2).
The parameter $\epsilon$ is the polarization of the virtual photon $\epsilon =
[4E_{beam}(E_{beam} - \nu) - Q^2]/[4E_{beam}(E_{beam} - \nu) + 2\nu^2 + Q^2]$. 
The term $\sigma_T$ represents the cross section for transverse photons,
$\sigma_L$ represents the cross section for longitudinal photons, $\sigma_{TT}
$ is the interference between the transverse amplitudes, and $\sigma_{LT}$ 
is the interference between transverse and longitudinal amplitudes.
The terms $\sigma_{TT}$ and $\sigma_{LT}$ approach zero
as $t \rightarrow 0$, and the terms $\sigma_L$ and $\sigma_{LT}$ vanish 
as $Q^2 \rightarrow 0$. In eq. (2), the structure functions $\sigma_T$ and 
$\sigma_{TT}$ can be further decomposed into two 
parts: $\sigma_\perp$ corresponds to incident photons polarized perpendicular 
to the hadronic plane, and $\sigma_\parallel$ corresponds to photons 
polarized parallel to the hadronic plane: $\sigma_T = (\sigma_
\parallel + \sigma_\perp)/2$, $\sigma_{TT} = (\sigma_\parallel - \sigma_\perp)
/2$ and $\sigma_{LT} = 2Re(A_LA^*_\parallel)$.\\
{\bf Transversely polarized target.} In eq. (2) we took explicitly into account 
the helicities of the virtual photon and ignored the helicities of ingoing and 
outgoing nucleon \cite{Bartl}. By taking into account the nucleon spin,
 $\sigma_\parallel =|A^N_\parallel|^2 + |A^F_\parallel|^2$,
 $\sigma_\perp = |A^N_\perp|^2 + |A^F_\perp|^2$,
 $\sigma_L =|A^N_L|^2 + |A^F_L|^2$, and 
 $\sigma_{LT} = 2Re(A_L^N A_\parallel^{N*} + A_L^F A_\parallel^{F*})$, where
N and F refer to nucleon flip and non-flip amplitudes, respectively.
In the $t$ channel, the contributions to $A^{N,F}_\perp$ come only from natural 
parity exchange, and contributions to $A^{N,F}_\parallel$ and $A^{N,F}_L$ come
only from unnatural parity exchange.
Using a transversely polarized target one obtains six more structure
functions which are the imaginary parts of products of non-flip and flip 
amplitudes $Im(A^N_i A^{F*}_j)$.

Determination of the pion form factor from electroproduction data requires the
extraction of that part of the cross section which contains the spin-flip
amplitudes, i.e. $|A^F_\parallel|^2$ and $|A^F_L|^2$. 

The ``inverse" reaction to $\gamma p \rightarrow \pi^+ n$ is the
$\pi^+ n \rightarrow \rho^o p$. Good quality data \cite{Grayer} exist 
only for the reaction $\pi^- p \rightarrow \rho^o n$ at $17.2$ 
$GeV$. The measured differential cross section, for the above reaction, as a 
function of $\sqrt{t}$ shows a behavior one would expect for one-pion exchange
mechanism (spin flip amplitude) which vanishes at $t = 0$. The most 
interesting observation for the  $\pi^- p \rightarrow \rho^o n$ reaction is
the presence of strong polarization effects \cite{Becker}, i.e. a large 
left-right polarized target asymmetry (presence of non-flip amplitudes)
in the low $t$ region. A sizable 
asymmetry was also observed \cite {Morehouse} in $\pi^+$ photoproduction from
a polarized target at 5 and 16 $GeV$. A typical value of the asymmetry is 
about $-0.5$ in both experiments. This is very surprising since,
according to general belief, this region should be dominated by one-pion 
exchange and should, therefore, show little or no polarization effects.   

\begin{flushleft}
{\bf 3. Simulation}
\end{flushleft}
\noindent
We have used the SDA Package \cite{Bogdan2} to simulate the 
$ep \rightarrow e \pi^+(K^+) n(\Lambda^o)$ reactions and to 
reconstruct the events accepted in the CLAS detector. 
In order to estimate rates we have used the following form for the differential 
cross sections \cite{Roberts}: 
\begin{eqnarray}
 { {d^2\sigma}\over{dQ^2dW} } = 
                { {\alpha W\sqrt{(W^2 + Q^2 -  m_p^2)^2 + 4m_p^2 Q^2}}
             \over {\pi(1 - \epsilon)(s - m_p^2)^2 Q^2} }  
 \cdot { {d\sigma_{\gamma_v p}(W,Q^2,\theta^*,\phi^*)} \over{d\Omega_{\pi,K}} }
\end{eqnarray}
where the first term is a flux $\Gamma$ of virtual photons, the second term, 
$\sigma_{\gamma_vp}$, represents the four structure functions as shown by eq. 
(2), and $s$ is the center-of-mass energy squared $s = m_p^2 + 2m_p E_{beam}$.\\ \\
In our simulation we have used the measured cross section $\sigma_{\gamma_v p}$
at $Q^2<1$ $GeV^2$ \cite{Brown} and extrapolated to higher $Q^2$ values with 
a simple pole form: $\sim (1 + Q^2/0.462)^{-2}$.

A sample of 1.6 million events was generated in the $Q^2$ range from 1.2 to 3.2
$GeV^2$ and in $W$ range from 2.05 to 2.15 $GeV$ for 5 beam energies: 4.0, 4.5, 
5.0, 5.5 and 6.0 $GeV$. Realistic trajectories of charged particles traversing 
the CLAS magnetic field were simulated, including multiple scattering and the 
drift cell spatial resolution of 250$\mu$m. 
For the purpose of the present study, both the scattered electron and the meson 
had to be detectable in the trigger scintillation counters and in all layers of 
the drift chambers in a given sector. Additionally, we required that the outgoing 
electron is within the acceptance of the \^{C}erenkov and Shower Counters. 
These requirements (acceptance) provide optimal trajectory reconstruction 
for both charged particles, and also a good missing mass resolution. 

Hereafter, we refer to the number of the generated events weighted by the cross 
section (see eq.(3)) at a given luminosity as the number of produced events, 
$N_{prod}$. A fraction of the $N_{prod}$ events that would have been 
accepted by the geometry of the CLAS detector is not observed because of 
reconstruction inefficiencies and various other losses like:  decaying pions 
(kaons), secondary interactions, radiative corrections, missing mass cut, etc.
These losses depend on the event kinematics and can be corrected for on an 
event-by-event bases \cite{Bogdan3}. In the present study, to account for these 
losses, we have introduced a constant global weight factor 
$w_g = N_{acc}/N_{obs} = 1.4$. 

During the reconstruction process we assumed that opposite sectors of 
the CLAS detector are not perfectly aligned, but are rotated relative to each 
other randomly by an angle of 1 mrad. We also have assumed that each nominal 
beam energy is randomly off by $0.1\%$. The $W$, $Q^2$, and $t$ regions were
chosen to obtain sufficient acceptance for at least 4 of the beam energies.\\
{\bf Rates.} In Table 1, using eq. (3), we show the expected rates of produced
$N_{prod}$ and fully reconstructed $N_{obs}$ events with a run of 100 
hours/$E_{beam}$ of the CLAS detector at a luminosity $L = 10^{34}cm^{-2}s^{-1}$ 
for $\Delta W = 2.05 - 2.15$ $GeV$ and 
$\Delta\Omega_\pi = 2\pi\int sin\theta^*_\pi d\theta^*_\pi = 0.377 sr$ 
($\theta^*_\pi < 20^o$). The analysis of the CLAS data taken at
a beam energy of 4 $GeV$ shows that the rates for semi-inclusive electroproduction
of charged pions are approximately larger by a factor of 15.

\begin{flushleft}
{\bf 4. Particle Identification}
\end{flushleft}
\noindent
To identify scattered electrons we first determine the clusters in the Shower
Counter. Negative tracks (potential electrons) which match these clusters are
selected. These tracks are then checked to determine whether the deposited
energy in the cluster agrees with the track momentum.
For our kinematical conditions, outgoing mesons have momenta ranging from $2.0$ 
to $3.5$ $GeV$, therefore the Time-of-Flight technique is not adequate. 
Hence, the outgoing mesons are identified using a missing mass technique.
Our preliminary analysis of 4 $GeV$ CLAS data indicate that the background under 
the missing mass peak of the neutron is only about a few percent.

\begin{flushleft}
{\bf 5. Analysis of Reconstructed Events}
\end{flushleft}
\noindent
From the reconstructed (observed) events $N_{obs}$, we extract the $Q^2$ and $t$ 
(or $\theta^*$) dependence of the structure functions: $\sigma_T$, $\sigma_L$, 
$\sigma_{TT}$ and $\sigma_{LT}$ in a model independent way. The procedure
consists of fitting the form of the differential cross section in eq. (3) to the
measured cross sections in a given kinematical bin $\Delta Q^2\Delta W\Delta t$.
The structure functions  $\sigma_T + \epsilon \sigma_L$, $\sigma_{TT}$ and
$\sigma_{LT}$ can be isolated experimentally using the $\phi$ dependence of the 
cross section at a given beam energy. A separation of $\sigma_T$ and 
$\sigma_L$ requires data measured at different beam energies. 

The measured cross section $(\partial\sigma_{\gamma_vp}/\partial\vec k)_m$ has 
been obtained from $N_{obs} \pm\sqrt{N_{obs}}$ in each bin $\Delta\vec k =
\Delta Q^2\Delta W\Delta\theta^*\Delta\phi^*$  in the following way:
\begin{eqnarray}
 \left({ {\partial\sigma_{\gamma_vp}} \over {\partial\vec k} }\right)_m = 
 { {N_{obs}(\Delta\vec k)\cdot w_g(E_b)_m}\over {A(\Delta\vec k,E_b)\cdot 
   L_m(E_b)\cdot F_m(\Delta W,\Delta Q^2,E_b)\cdot \Delta\vec k_m} }  
\end{eqnarray}

The average acceptance $A_m(\Delta\vec k,E_b)$ in each bin $\Delta\vec k$ and 
at each beam energy has been determined by generating at random 400 events/bin. 
These events were transformed into the lab system with a random rotation of the 
electron scattering plane around the beam direction, and were checked to
determine if the outgoing electron and meson both lie within the
geometry of the CLAS detector \cite{Bogdan3}. In Table 2, as an example, we show 
the acceptance as a function of the c.m. angles ($\theta^*_\pi$,$\phi^*_\pi$)
for $\Delta W = (2.05 - 2.15)$ $GeV$, $\Delta Q^2 = (2.0 - 2.2)$ $GeV^2$ and 
$E_{beam} = 4.5$ $GeV$. The measured flux $F_m(\Delta W,\Delta Q^2,E_b)$, 
in eq. (4), was obtained by averaging over all events observed in a bin
$\Delta W\Delta Q^2$ and for given beam energy.     
In Fig. 1 we show, as an example, the derived differential cross sections as a
function of $\phi^*$ and $E_{beam}$ for one bin $k$ of size $\Delta W\Delta 
Q^2\Delta t$. 

To extract the four structure functions from the measured cross sections in a bin
$k$, as shown in Fig. 1, we used the following functional form in the $\chi^2$ 
minimization procedure:
\begin{eqnarray}
 f(\phi^*)_k = \sum_i \left( P_1+P_2\cdot\epsilon_i+P_3\cdot\epsilon_i \cdot
  cos2\phi^* +P_4\cdot\sqrt{\epsilon_i(\epsilon_i+1)/2}\cdot cos\phi^*\right)  
\end{eqnarray}
where the index $i$ runs over the different beam energies.

The four parameters ($P_1,P_2,P_3$, and $P_4$) are determined from 
the fit. The virtual photon polarization $\epsilon_i$ was evaluated at the 
center of the $\Delta W$ and $\Delta Q^2$ bins. 
Data with $A(\Delta\vec k,E_b) < 2\%$ were excluded from the fit, see Table 1.
The $Q^2$ dependence of the four structure 
functions and their statistical errors, derived from the fits, are summarized in 
Fig. 2 for pions only (see full circles with error bars). The dashed or solid
curves in Fig. 2 represent the input to the Monte Carlo simulation. The full 
squares and the stars in Fig. 2 represent the currently available data 
\cite{Bebek1}. The fitting procedure also includes the expected systematic errors
in determining the luminosity $L_m(E_b)$ of $\leq 1\%$ and a global correction 
$w_g(E_b)_m$ of $\leq 2\%$; i.e., uncorrelated systematic 
errors at each beam energy setting. In addition to the above, we introduced a 
correlated systematic error (common to all beam energies) of about $3\%$.
The succesfull $\sigma_L$ and $\sigma_T$ separation was possible, in the above 
kinematical regions, once sufficient data were collected for at least 4 of 
the 5 beam energies.

\newpage
\begin{flushleft}
{\bf 6. Summary}
\end{flushleft} 
\noindent
A challenging problem in particle physics is to understand the transition from 
the ``current" quarks and gluons appearing in the QCD Lagrangian to the degrees 
of freedom of low-energy QCD.
One could take the approach that anything that can be calculated by pQCD can be
called a hard process. All the rest would be soft. Soft interactions are usually
understood as the interactions of hadrons at a relatively small scale (low $Q^2$
in $ep$ interactions or low $p_T$ in hadron-hadron interactions). 
The problem, however, is that what we calculate and what we measure are not the 
same. Soft interactions are not easily disentangled from hard ones \cite{Levy}.\\ 
That said, let us summarize here what we can measure in a model independent way.
\\ \\
{\bf Stage 0.} Analyze the existing data taken at a beam energy of 4 $GeV$.
Extract the $Q^2$ and $t$ dependences of the structure functions: $\sigma_T + 
\epsilon \sigma_L$, $\sigma_{TT}$, and $\sigma_{LT}$.\\
{\bf Stage 1.} Take new data at beam energies: 4.5, 5.0, 5.5 and 6.0 $GeV$,
and now disentangle the $\sigma_L$ and $\sigma_T$ cross sections.\\
{\bf Stage 2.} Measure, for the first time, asymmetries for exclusive and 
semi-inclusive electroproduction on transversely polarized target.
Technically, it is a challenging task to operate CLAS with the strong transverse 
magnetic field of a polarized target.

\begin{flushleft}
{\bf Conclusion}
\end{flushleft} 
\noindent
We have shown that using the CLAS spectrometer at JLAB and with beam energies
between 4 and 6 $GeV$, we can obtain good quality electroproduction data that 
will improve our understanding of the nucleon structure as well as the hadronic
properties of the photon. We emphasize the importance of studying both the $Q^2$
and $t$ dependencies of the various structure functions for $\pi^+$, $\pi^o$
and $K^+$ exclusive electroproduction.  
In order to access higher $W$ and $Q^2$ at higher beam energies, we need to
design a new large acceptance detector in the forward direction.\\

\begin{flushleft}
{\bf Acknowledgments}
\end{flushleft} 
\noindent 
We have benefited from discussions with M. Guidal, G. Piller, A.V. Radyushkin,
and A. Yegneswaran.

\newpage
\begin{figure}[p]
\epsfxsize=6.0in
\begin{center}
\epsfbox{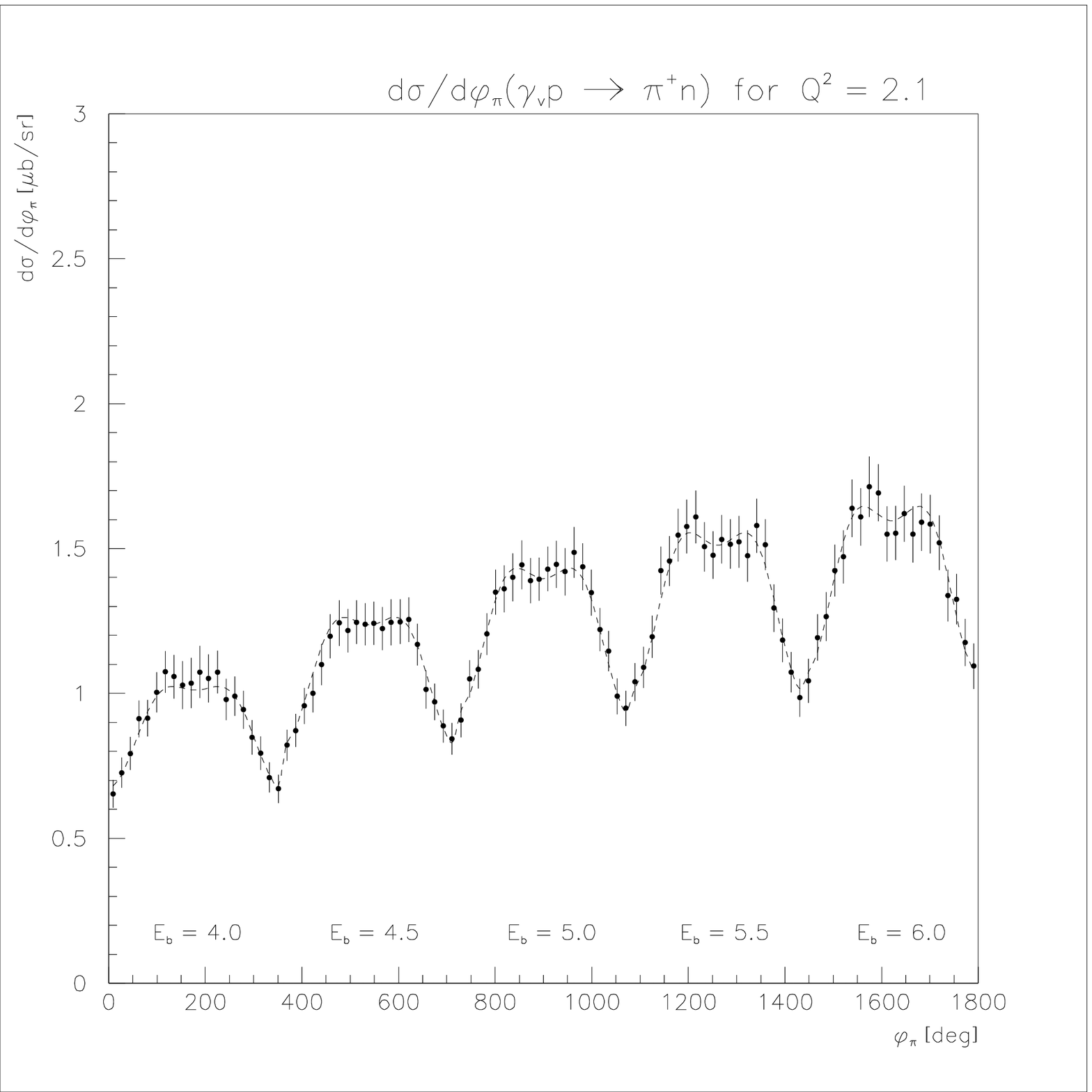}
\caption [An example of a differential cross section.]
         {An example of the derived differential cross section as a function
          of $\phi^*$ and $E_{beam}$ for $Q^2 = 2.1$ $GeV^2$.}
\end{center}
\end{figure}

\newpage
\begin{figure}[p]
\epsfxsize=6.0in
\begin{center}
\epsfbox{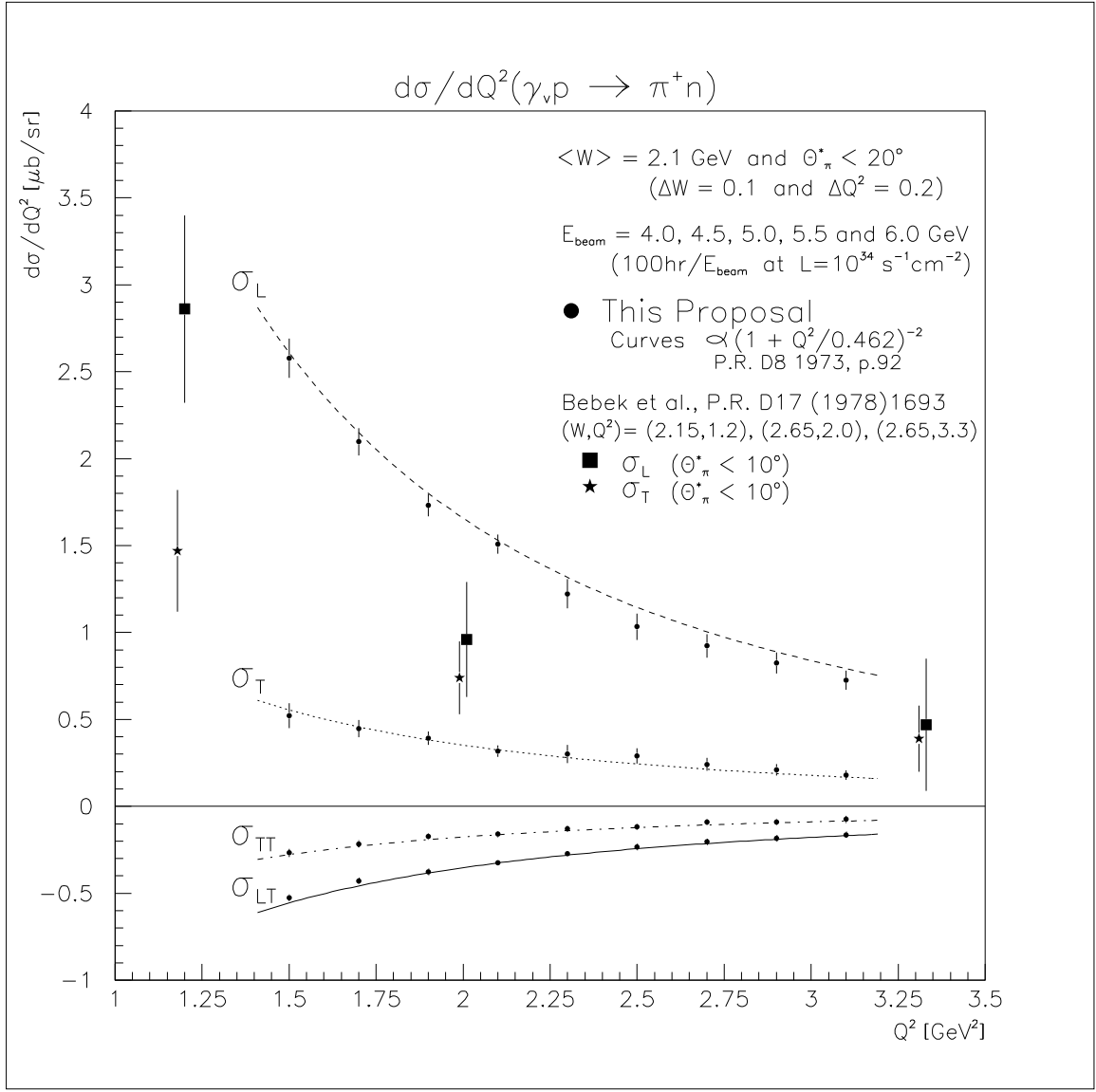}
\caption [$Q^2$ dependence of structure functions for pions.]
       {The $Q^2$ dependence of the four structure functions 
       and their statistical errors, derived from fits, for pions.
       Existing data \cite{Bebek1}: $\sigma_L$ ($\Box$) and $\sigma_T$ ($\star$)} 
\end{center}
\end{figure}

\newpage
\begin{table}[p]
\caption [Expected rates.]
 {Expected rates of produced ($N_{prod}$/100hr) and reconstructed 
  ($N_{obs}$/100hr) $\pi^+$ mesons using the CLAS detector at a luminosity of 
  $10^{34}cm^{-2}s^{-1}$ for $\Delta W = (2.05 - 2.15)$ $GeV$ and 
  $\Delta\Omega_\pi = 0.377 sr$ ($\theta^*_\pi < 20^o$).}
\begin{center}
  \begin{tabular} {|c||r|r|r|r|r|} \hline
   & $E_b=4.0$& $E_b=4.5$& $E_b=5.0$& $E_b=5.5$& $E_b=6.0$ \\ \hline
   $\Delta Q^2$& $N_{prod}$& $N_{prod}$& $N_{prod}$& $N_{prod}$& $N_{prod}$ \\
   & $N_{obs}$& $N_{obs}$& $N_{obs}$& $N_{obs}$& $N_{obs}$ \\ 
   \hline\hline
   1.4 - 1.6  & 36349 & 43455 & 49217 & 53913 &       \\
              & 13226 & 13547 & 11511 &  6677 &       \\ \hline 
   1.6 - 1.8  & 23494 & 28537 & 32698 & 36107 & 38980 \\
              &  8946 & 10395 & 10073 &  8091 &  5059 \\ \hline 
   1.8 - 2.0  & 15564 & 19372 & 22451 & 25049 & 27130 \\
              &  5968 &  7566 &  7755 &  7283 &  5957 \\ \hline 
   2.0 - 2.2  & 10527 & 13480 & 15805 & 17756 & 19441 \\
              &  3918 &  5440 &  5999 &  5975 &  5343 \\ \hline 
   2.2 - 2.4  &  7262 &  9545 & 11399 & 12926 & 14175 \\
              &  2552 &  3987 &  4533 &  4692 &  4511 \\ \hline 
   2.4 - 2.6  &       &  6881 &  8325 &  9571 & 10511 \\
              &       &  2855 &  3494 &  3738 &  3678 \\ \hline 
   2.6 - 2.8  &       &  5011 &  6196 &  7182 &  8012 \\
              &       &  2017 &  2645 &  2942 &  3016 \\ \hline 
   2.8 - 3.0  &       &  3682 &  4670 &  5472 &  6144 \\
              &       &  1419 &  1994 &  2311 &  2457 \\ \hline 
   3.0 - 3.2  &       &  2734 &  3541 &  4219 &  4779 \\
              &       &   974 &  1487 &  1792 &  1960 \\ \hline 
  \end{tabular}
\end{center}
\end{table}

\newpage
\begin{table}[p]
\caption [An example of acceptance.]
 {An example of the acceptance as a function of the c.m. angles 
 ($\theta^*_\pi$,$\phi^*_\pi$) for $\Delta W = (2.05 - 2.15)$ $GeV$, 
 $\Delta Q^2 = (2.0 - 2.2)$ $GeV^2$ and $E_{beam} = 4.5$ $GeV$.}
\begin{center}
  \begin{tabular} {|c||c|c|c|c|} \hline
   &$\Delta\theta^*_\pi$ $[^o]$ &$\Delta\theta^*_\pi$ $ [^o]$ 
   &$\Delta\theta^*_\pi$ $ [^o]$ &$\Delta\theta^*_\pi$ $ [^o]$ \\
   $\Delta\phi^*_\pi$ $ [^o]$& 0 - 5  & 5 - 10 & 10 - 15& 15 - 20\\ \hline\hline
    0 - 18   & 0.72  & 0.70  & 0.67  & 0.68  \\ \hline
   18 - 36   & 0.67  & 0.69  & 0.64  & 0.63  \\ \hline
   36 - 54   & 0.68  & 0.66  & 0.57  & 0.51  \\ \hline
   54 - 72   & 0.64  & 0.54  & 0.53  & 0.48  \\ \hline
   72 - 90   & 0.64  & 0.59  & 0.48  & 0.45  \\ \hline
   90 - 108  & 0.64  & 0.55  & 0.45  & 0.39  \\ \hline
   108 - 126 & 0.68  & 0.52  & 0.43  & 0.39  \\ \hline
   125 - 144 & 0.70  & 0.58  & 0.46  & 0.41  \\ \hline
   144 - 162 & 0.64  & 0.60  & 0.52  & 0.44  \\ \hline
   162 - 180 & 0.63  & 0.66  & 0.60  & 0.42  \\ \hline
   180 - 198 & 0.67  & 0.61  & 0.57  & 0.41  \\ \hline
   198 - 216 & 0.66  & 0.60  & 0.56  & 0.41  \\ \hline
   216 - 234 & 0.64  & 0.55  & 0.44  & 0.34  \\ \hline
   234 - 252 & 0.64  & 0.55  & 0.49  & 0.36  \\ \hline
   252 - 270 & 0.63  & 0.52  & 0.50  & 0.45  \\ \hline
   270 - 288 & 0.67  & 0.59  & 0.52  & 0.43  \\ \hline
   288 - 306 & 0.67  & 0.61  & 0.54  & 0.46  \\ \hline
   306 - 324 & 0.69  & 0.63  & 0.53  & 0.56  \\ \hline
   324 - 342 & 0.73  & 0.69  & 0.66  & 0.59  \\ \hline
   342 - 360 & 0.68  & 0.71  & 0.66  & 0.70  \\ \hline
  \end{tabular}
\end{center}
\end{table}

\end{document}